**Stanislav Levytskyi (Ukraine),**
**Oleksandr Gneushev (Ukraine),**
**Vasyl Makhlinets (Ukraine)**

# THE APPROACH TO MODELING THE VALUE OF STATISTICAL LIFE USING AVERAGE PER CAPITA INCOME

**Abstract**

The problem of determining the value of statistical life in Ukraine in order to find ways to improve it is an urgent one now. The current level of value is analyzed, which is a direct consequence of the poor quality of life of a citizen, hence his low level. The description of the basic theoretical and methodological approaches to the estimation of the cost of human life is given. Based on the analysis, a number of hypotheses have been advanced about the use of statistical calculations to achieve the modeling objectives. Model calculations are based on the example of Zaporozhye Oblast statistics for 2018–2019.

The article elaborates the approach to the estimation of the economic equivalent of the cost of living on the basis of demographic indicators and average per capita income, and also analyzes the posibilities of their application in the realities of the national economy. Using Statistica, the regession equation parameters were determined for statistical data of population distribution of Zaporizhzhia region by age groups for 2018. The calculation parameters were also found using the Excel office application, using the Solution Finder option to justify the quantitative range of metric values. It is proved that the proposed approach to modeling and calculations are simpler and more efficient than the calculation methods proposed earlier. The study concluded that the value of statistical life in Ukraine is significantly undervalued.

| **Keywords:** | economic equivalent of value of life, cost of statistical life, average per capita income, modeling, demographic statistics |
|---|---|
| **JEL Classification:** | C19, J17 |

**С. І. Левицький (Україна),**
**О. М. Гнєушев (Україна),**
**В. М. Махлинець (Україна)**

# ПІДХІД ДО МОДЕЛЮВАННЯ ВАРТОСТІ СТАТИСТИЧНОГО ЖИТТЯ З ВИКОРИСТАННЯМ СЕРЕДНЬОДУШОВОГО ДОХОДУ

**Анотація**

У теперішній час актуальною є проблема визначення оцінки вартості статистичного життя в Україні з метою пошуку шляхів її підвищення. Проаналізовано поточний рівень вартості, що є прямим наслідком незадовільної якості життя громадянина, отже, його низького рівня. Наведено характеристику основних теоретико-методологічних підходів до оцінки вартості життя людини. На базі проведеного аналізу висунуто низку гіпотез про використання статистичних розрахунків для досягнення цілей моделювання. Модельні розрахунки наведено на прикладі статистичних даних Запорізької області за 2018–2019 рр.

В статті розроблено підхід до оцінки економічного еквіваленту вартості життя на основі демографічних показників та середньодушового доходу, а також проведено аналіз можливостей їх застосування в реаліях національної економіки. За допомогою ППП «Statistica» для статистичних даних розподілу населення Запорізької області по віковим групам за 2018 рік визначено параметри регресійного рівняння. Також знайдено розрахункові параметри за допомогою офісного додатку Excel, застосовуючи опцію «Пошук рішення» для обґрунтування пропозицій кількісного діапазону показників. Доведено, що запропонований підхід до моделювання та розрахунки є більш простими та ефективними, ніж методики розрахунків, запропонованих раніше. За результатами дослідження зроблено висновок, що цінність статистичного життя в Україні істотно недооцінена.

| | |
|---|---|
| **Ключові слова** | економічний еквівалент вартості життя, вартість статистичного життя, середньодушовий дохід, моделювання, демографічна статистика |
| **Класифікація JEL** | C19, J17 |

# ВСТУП

Матеріальна відповідальність за життя людини - це суттєвий важіль стимулювання зусиль з модернізації систем забезпечення безпеки людей у різних вимірах, що, безумовно, у кінцевому рахунку веде до підвищення якості життя всього народу. Економічна оцінка вартості статистичного життя, відображаючи рівень його якості є показником соціально-економічного розвитку суспільства, а також індикатором діагностики виникнення кризових явищ. Низка вартість життя громадянина є наслідком його незадовільної якості, отже, низького рівня життя в країні. Цю обставину підтверджує більшість наукових досліджень: чим багатша країна, тим вище вона оцінює життя своїх громадян. Тому подібні дослідження сприятимуть як глибшому розумінню природи соціально-економічних явищ в національній економіці, так і пошуку раціональних способів покращення життя населення, що є у теперішній час особливо актуальним для України, зокрема, для Запорізької області.

# 1. ЛІТЕРАТУРНИЙ ОГЛЯД

На теперішній час існують різні погляди на розрахунки показників вартості життя в Україні, зокрема, на розрахунок вартості статистичного життя людини. Це призводить до застосування різних методичних підходів та некоректного їх використання [4, 15]. Жоден з цих підходів не може вважатися абсолютно точним. Під егідою Європейської обсерваторії по системам і політиці охорони здоров'я вчені досліджували характер залежності між здоров'ям населення і динамікою розвитку ВВП держави [13]. Результати цих досліджень в різних країнах світу переконливо свідчать, що стан здоров'я населення є потужним фактором економічного зростання, оскільки приріст ВВП безпосередньо залежить від рівня смертності населення [12]. Негативною тенденцією є катастрофічно високий рівень смертності людей працездатного віку, серед яких біля 80% - це чоловіки. А в цілому смертність населення в працездатному віці по Україні в 3.5 рази вище, ніж в європейських країнах. Висновки вказаних дослідників говорять про те, що високий рівень смертності членів суспільства істотно гальмує економічне зростання держави. Показник «Втрачені роки потенційного життя» (в англомовній літературі - Potential Years of Life Lost, PYLL) є одним із сучасних інструментів оцінки втрат здоров'я населення, що дозволяє оцінювати ці втрати в економічному аспекті та відноситься до числа широко розповсюджених в світі для оцінки здоров'я і благополуччя населення. Ним користуються, зокрема, Світовий банк, Організація економічного співробітництва і розвитку, Всесвітня організація охорони здоров'я і Євросоюз [9]. При розрахунку втрачених років потенційного життя (ВПРЖ) визначається число років, які не доживає популяція до деякого нормативного віку. За узгодженою більшістю експертів думкою, такий нормативний вік дорівнює 70 рокам, хоча в ряді країн використовуються і інші вікові межі, наприклад, 65 або 75 років [11]. В Україні такий нормативний вік дорівнює 65 рокам [5]. По статистичним даним Запорізької області за 2017 рік і по розрахунках ВПРЖ складає 110,000 років, а втрата ВРП в області склала 8.6% [11]. Таким чином, оцінка вартості статистичного життя людини в Україні має відбуватися лише з урахуванням поточної економічної ситуації у національній економіці.

На даний час в Україні показник здоров'я населення не враховують при розрахунку ВВП в силу того, що здоров'я не є ринковим товаром, а значить, не має ринкової вартості. Висока ціна економічного еквіваленту людського життя, закріплена законодавчо, стимулює державні заходи у вживанні заходів по скороченню смертності, підвищення дорожньої і виробничої безпеки, у веденні більш ефективної політики в галузі охорони здоров'я, тощо [6-8]. Якщо прийняти як еквівалент вартості року життя українця величину середньорічної заробітної плати в 2013 році, то ціна втрачених років потенційного життя становить 118 млрд грн [2]. Економічна оцінка втрат внаслідок передчасної смертності в Україні має вигляд, наведений у Таблиці 1 по розрахунках у [13].

**Таблиця 1.** Економічна оцінка втрат внаслідок передчасної смертності в Україні (2006, 2011, 2013 рр.)

*Джерело: Складено авторами за [13].*

| Показник | 2006 | 2011 | 2013 |
|---|---|---|---|
| Кількість втрачених років потенційного життя, млн людино-років | 4.165 | 3.134 | 3.031 |
| Оцінка втрат у вигляді недоотриманого ВВП, млрд грн | 47.9 | 90.4 | 97.1 |
| Частка від ВВП, % | 8.8 | 6.7 | 6.7 |

Основні підходи, які можуть бути використані при оцінці вартості людського життя в Україні, наведено у Таблиці 2.

**Таблиця 2.** Основні теоретико-методологічні підходи оцінки вартості людського життя

*Джерело: Розробка авторів за джерелом [12].*

| Найменування основних теоретико-методологічних підходів до оцінки вартості життя людини | Коротка характеристика основних теоретико-методологічних підходів до оцінки вартості життя людини |
|---|---|
| Підхід з позицій нормативно встановленого відшкодування у зв'язку із загибеллю людини і виконання рішення суду | Ґрунтується на законодавчому рішенні суду, що визначив максимальний розмір компенсації по відшкодуванню збитку здоров'ю і життю постраждалої людини від дії фізичної або юридичної особи в результаті надзвичайної ситуації (НС) |
| Підхід з позиції корисності людини для суспільства | Ґрунтується на теорії корисності і спрямований на розрахунок економічної або громадської корисності людини для суспільства при настанні тимчасової або стійкої втрати людиною працездатності або його передчасної смерті; виражається через показник невиробленого ВВП із-за загибелі людини в результаті НС |
| Підхід з точки зору корисності людини для домогосподарства | Ґрунтується на офіційних економіко-демографічних показниках і дозволяє визначити вартість життя як різницю накопичених і спожитих людиною матеріальних благ і послуг на підставі майбутнього можливого заробітку людини |
| Соціологічний підхід | Ґрунтується на соціологічних опитуваннях різних соціальних груп населення країни (регіону, міста) і дозволяє визначити економічний еквівалент «вартості» життя середньої людини як величини «достатнього» і «справедливого» відшкодування у зв'язку із загибеллю людини в результаті НС - на думку респондентів |
| Підхід з позиції оцінки ризиків | Ґрунтується на економічній оцінці ризику нанесення збитку здоров'ю і життю людини в умовах НС за допомогою визначення розміру грошового еквіваленту, який суспільство готове заплатити за зменшення, уникнення або відвертання дії НС або використати в якості компенсації людині за понесені втрати |
| Підхід з позиції готовності фізичних осіб платити за усунення ризику смерті | Ґрунтується на гіпотетичній готовності фізичних осіб платити за усунення ризику смерті від конкретних керованих зовнішніх чинників НС |
| Підхід з позицій страхування вартості життя для окремих груп населення | Ґрунтується на визначенні страховими компаніями розміру страхових сум страхових внесків і компенсації вартості життю для окремих соціальних і професійних груп населення |
| Підхід з позицій вартості медичних послуг, що забезпечують зниження ризику передчасної смерті | Ґрунтується на розрахунках реальних і прогнозованих максимальних витрат суспільства на надання медичних послуг, що забезпечують зниження ризику передчасної смерті людини |
| Метод особистого капіталу | Ґрунтується на оцінці заробітку (сукупного доходу) людини протягом усього життя |

Величина економічного еквіваленту вартості життя (ЕЕВЖ) людини (наприклад, компенсації збитку в надзвичайних ситуаціях (НС) у зв'язку із загибеллю людини) повинна відповідати наступним основним вимогам [12]:

1) відповідність суб'єктивній вимозі «справедливості» - значна більшість дорослого населення країни повинна розглядати відшкодування у зв'язку із загибеллю людини при НС різного характеру як достатнє для компенсації понесеного збитку;
2) достатність відшкодування для компенсації сумарного матеріального збитку, реально понесеного домогосподарством у зв'язку із загибеллю людини в результаті НС;
3) достатність компенсації для відшкодування морального збитку (моральних страждань), понесених близькими в результаті загибелі людини із-за НС.

## 2. МЕТА ДОСЛІДЖЕННЯ

Розробка підходів до оцінки економічного еквіваленту вартості життя на основі демографічних показників та середньодушового доходу, а також аналіз можливостей їх застосування в реаліях національної економіки з обґрунтуванням пропозицій щодо кількісного діапазону значень цього показника на прикладі Запорізької області є метою статті.

## 3. МЕТОДИ ДОСЛІДЖЕННЯ

В результаті аналізу було встановлено, що людина своєю економічною і фізично небезпечною (безпечною) для свого життя поведінкою оцінює своє життя таким [1]:

$$E(T_ж) = \frac{Д_{с2}}{P_y}, \qquad (1)$$

де $E(T_ж)$ - економічний еквівалент вартості життя середньостатистичної людини в середньому віці, $Д_{с2}$ - середньодушовий використовуваний грошовий дохід у вигляді заробітної плати, доходів від підприємницької діяльності, пенсій, стипендій, соціальних трансфертів, доходів від власності, дивідендів, процентів та інших доходів за вирахуванням обов'язкових платежів: податків, квартплати, вартості комунальних послуг та інших фінансових зобов'язань, $P_y$ - фоновий ризик смерті людей (ймовірність померти від будь-якої причини смерті), у демографії цей показник називається коефіцієнтом смертності $K_с$ з урахуванням всіх причин смерті людей, $T_ж$ - середній вік людей, які проживають у країні,

$$P_y = K_с = \frac{\text{число людей, які померли в країні за 1 рік зі всіх причин}}{\text{середньорічна чисельність населення країни}}. \qquad (2)$$

Середньорічна чисельність населення – розрахована як середня арифметична з чисельності на початок і кінець календарного року за даними Державного комітету статистики України. Середній вік людей $T_ж$, які проживають у країні, обчислюється на підставі таблиць розподілу населення кожної країни по віковим групам, вказаних в статистичних даних [5].

Розподіл осіб за однорічними віковими групами є інтервальним, оскільки на момент спостереження особа прожила не тільки число зазначених років, але й ще певну кількість днів, місяців (час є безперервним). Тобто запис «$x$ років» відповідає інтервалу «з $x$ до $x+1$ років». Тому при розрахунку середнього віку, що живуть, береться середина інтервалу $[x ; x+1)$ років:

$$\overline{x} = \frac{x + x + 1}{2}. \qquad (3)$$

Таким чином середній вік людей $T_ж$, що живуть, обчислюється за формулою:

$$T_ж = \frac{\sum_{x=0}^{100}(S_x \times \overline{x})}{\sum_{x=0}^{100} S_x}, \qquad (4)$$

де $S_x$ - середня чисельність населення віку $x$.

Середній вік людей $T_ж$, що живуть, може бути виражений через закон $f(t_ж)$ Вейбулла-Гнеденко щільності розподілу ймовірностей віку $t_ж$ людей, що живуть [1]:

$$T_ж = \int_0^\infty t_ж f(t_ж) dt = a \cdot \Gamma(1 + \frac{1}{b}) + c, \qquad (5)$$

де $f(t_ж) = \dfrac{b}{a}\left(\dfrac{t_ж-c}{a}\right)^{b-1} \exp\left[-\left(\dfrac{t_ж-c}{a}\right)^b\right]$, $a, b, c$ - параметри щільності розподілу ймовірності віку $t_ж$ людей, що живуть, $a$ - параметр масштабу, $b$ - параметр форми, $c$ - параметр зсуву, $\Gamma$ - гамма-функція:

$$\Gamma(y) = \int_0^\infty x^{y-1} e^{-x} dx. \qquad (6)$$

Функція розподілу $F(x)=P(X<x)$ знаходиться за формулою:

$$F(t_ж) = 1 - \exp\left[-\left(\dfrac{t_ж-c}{a}\right)^b\right], t_ж \geq c, a, b > 0, \qquad (7)$$

де $F(t_ж)$ при $t_ж < c$.

ЕЕВЖ новонародженої людини:

$$E_o = \dfrac{E(T_ж)}{\exp\left[-\left(\dfrac{T_ж-c}{a}\right)^b\right]}. \qquad (8)$$

ЕЕВЖ людини віку $t_ж$ років:

$$E(t_ж) = E_o \exp\left[-\left(\dfrac{t_ж-c}{a}\right)^b\right]. \qquad (9)$$

Параметр $c=0$, так як вікова структура населення починається з $t_ж=0$.

Параметри $a, b$ можуть бути обчислені за статистичними даними про вікову структуру населення зведенням функції розподілу до лінійної функції за допомогою подвійного логарифмування [10]:

$$Y = AX + B, \qquad (10)$$

де $Y=\ln(-\ln(1-F))$, $X=\ln x$.

При цьому параметр $b=A$, $a=\exp(-B/b)$.

Для визначення ЕЕВЖ на основі теорії корисності за допомогою середньодушового річного доходу найчастіше застосовуються наступні підходи [3, 9, 11, 14]: для оцінки економічної корисності людини використовується значення середньодушового річного доходу; для оцінки економічної корисності людини використовується значення показника валового внутрішнього продукту на душу населення.

При першому підході вводиться гіпотеза, відповідно до якої економічна корисність людини для суспільства покладається рівною доходу, який він забезпечує для себе. При такому підході середньорічний дохід на людину є кількісна характеристика громадської корисності середньостатистичної людини.

При умові, що середньодушовий річний дохід і ставка дисконтування залишаються постійними оцінка економічної корисності середньостатистичної людини знаходиться через суму дисконтного душового доходу за період очікуваної тривалості майбутнього життя за наступною формулою:

$$ЕЕВЖ = Д_{сер} \cdot \int_0^{t_0} e^{-Et} dt \qquad (11)$$

де $Д_{сер}$ - середньодушовий річний дохід, $E$ – ставка дисконтування, $t_0 = e^0_{t_ж} - T_ж$ - період очікуваної тривалості майбутнього життя, $T_ж$ - середній вік людей, що живуть; $e^0_{t_ж}$ - середня очікувана тривалість життя при народженні.

Ставку дисконтування можна визначити за фактичною річною банківською відсотковою ставкою *i* таким чином:

$$E = ln(1+i). \qquad (12)$$

Оцінку статистичної вартості життя людини в Україні може бути здійснено в межах розрахунків за допомогою ВВП на душу населення, що дасть більш наближені до реальності результати дослідження на макрорівні. Гіпотеза про зв'язок збитку внаслідок передчасної смерті людини та середньодушового ВВП, скоригованого на ставку дисконту, підтверджується у публікаціях [1, 2, 3].

## 4. РЕЗУЛЬТАТИ

Для Запорізької області показник $Д_{с2}$ (обидві статі) обчислений таким чином. У 2018 році у складі грошових витрат населення обов'язкові податки на доходи, майно і інші трансферти склали 12.702 млн грн при чисельності населення 1,713.715 чоловік, що на душу населення складає: $\frac{12,702,000,000}{1,713.715} = 7.412$ грн.

Річні доходи на душу населення склали 67.982 грн. Тоді середньодушовий грошовий річний дохід, що розраховується, складає: $Д_{с2}$=67.982-7.412=60.570 грн.

Фоновий ризик смерті людей $P_y=K_c$ (коефіцієнт смертності з урахуванням всіх причин смерті людей) і очікувана тривалість життя для Запорізької області за 2018 р. (обидві статі) розраховані за допомогою таблиць смертності і середньої очікуваної тривалості майбутнього життя:

$$P_y = K_c = \frac{27.871}{1,713.715} = 0.016263. \qquad (13)$$

Таким чином, в Запорізькій області (обидві статі) за 2018 р. по формулі (4) визначаємо:

$$T_ж = \frac{72,603.285}{1,713.715} = 42.4 \text{ років.} \qquad (14)$$

Підставляючи у формулу (1) відповідні дані, обчислюється економічний еквівалент вартості життя середньостатистичної людини у віці $T_ж$=42.4 років в Запорізькій області за 2018 р. (обидві статі):

$$E(T_ж) = \frac{Д_{с2}}{K_c} = \frac{60.570}{0.016263} = 3,724.291 \text{ (грн).} \qquad (15)$$

Методика оцінки ЕЕВЖ ($E(t_ж)$) для довільного віку $t_ж$ пропонується в такому вигляді: $E(t_ж)$ буде більше (менше) в стільки разів, в скільки його очікувана тривалість життя буде більше (менше) очікуваної тривалості життя у віці $E(T_ж)$.

$E(t_ж)$ для віку $t_ж$ по статистичним даним розраховується за допомогою формули:

$$E(t_ж) = \frac{e_{t_ж}}{e_{T_ж}} \cdot E(T_ж), \qquad (16)$$

де $e_{t_ж}$ - середня очікувана тривалість життя для осіб, яким виповнилося $t_ж$ років, $e_{T_ж}$ - середня очікувана тривалість життя для осіб, яким виповнилося $T_ж$ років.

Середня очікувана тривалість життя для осіб (обидві статі), яким виповнилося $t_ж$ років для Запорізької області (2019 р.) в авторських розрахунках представлено у Додатку 1.

Згідно табличних даних маємо: $e_{T_ж} = e_{42.4} \approx 31.32$, $e_0 \approx 70.89$.

$$E(t_ж = 0) = E_0 = \frac{70.89}{31.32} \cdot 3,724.291 = 8,429.309 \text{ (грн).} \qquad (17)$$

Оцінки ЕЕВЖ для населення Запорізької області (обидві статі, 2018 р.) по формулі (16) і статистичним даним подані в Таблиці 3.

**Таблиця 3.** ЕЕВЖ для Запорізької області (обидві статі, 2018 р.)

Джерело: Розробка авторів за джерелом [12].

| Вік (років) | ЕЕВЖ (грн) | Вік (років) | ЕЕВЖ (грн) | Вік (років) | ЕЕВЖ (грн) | Вік (років) | ЕЕВЖ (грн) |
|---|---|---|---|---|---|---|---|
| 0 | 8,429.309 | 26 | 5,491.784 | 51 | 2,879.661 | 76 | 1,005.007 |
| 1 | 8,377.488 | 27 | 5,379.827 | 52 | 2,786.770 | 77 | 952.228 |
| 2 | 8,260.065 | 28 | 5,268.285 | 53 | 2,695.102 | 78 | 901.257 |
| 3 | 8,142.740 | 29 | 5,157.180 | 54 | 2,604.697 | 79 | 852.087 |
| 4 | 8,025.521 | 30 | 5,046.536 | 55 | 2,515.594 | 80 | 804.703 |
| 5 | 7,908.414 | 31 | 4,936.374 | 56 | 2,427.832 | 81 | 759.091 |
| 6 | 7,791.426 | 32 | 4,826.722 | 57 | 2,341.450 | 82 | 715.231 |
| 7 | 7,674.565 | 33 | 4,717.603 | 58 | 2,256.486 | 83 | 673.099 |
| 8 | 7,557.838 | 34 | 4,609.046 | 59 | 2,172.976 | 84 | 632.668 |
| 9 | 7,441.253 | 35 | 4,501.077 | 60 | 2,090.957 | 85 | 593.906 |
| 10 | 7,324.820 | 36 | 4,393.725 | 61 | 2,010.465 | 86 | 556.779 |
| 11 | 7,208.547 | 37 | 4,287.020 | 62 | 1,931.533 | 87 | 521.246 |
| 12 | 7,092.445 | 38 | 4,180.992 | 63 | 1,854.194 | 88 | 487.258 |
| 13 | 6,976.524 | 39 | 4,075.672 | 64 | 1,778.478 | 89 | 454.762 |
| 14 | 6,860.794 | 40 | 3,971.093 | 65 | 1,704.416 | 90 | 423.691 |
| 15 | 6,745.267 | 41 | 3,867.289 | 66 | 1,632.034 | 91 | 393.965 |
| 16 | 6,629.955 | 42 | 3,764.292 | 67 | 1,561.358 | 92 | 365.477 |
| 17 | 6,514.870 | 43 | 3,662.138 | 68 | 1,492.411 | 93 | 338.087 |
| 18 | 6,400.027 | 44 | 3,560.862 | 69 | 1,425.213 | 94 | 311.589 |
| 19 | 6,285.438 | 45 | 3,460.502 | 70 | 1,359.784 | 95 | 285.674 |
| 20 | 6,171.120 | 46 | 3,361.093 | 71 | 1,296.138 | 96 | 259.841 |
| 21 | 6,057.086 | 47 | 3,262.674 | 72 | 1,234.288 | 97 | 233.239 |
| 22 | 5,943.353 | 48 | 3,165.282 | 73 | 1,174.245 | 98 | 204.330 |
| 23 | 5,829.938 | 49 | 3,068.957 | 74 | 1,116.015 | 99 | 170.186 |
| 24 | 5,716.860 | 50 | 2,973.737 | 75 | 1,059.602 | 100 | 124.893 |
| 25 | 5,604.135 | – | – | – | – | – | – |

Щільність розподілу ймовірності $f(t_ж)$ віку $t_ж$ людей, що живуть по закону Вейбулла-Гнеденко вікових груп населення Запорізької області за 2018 рік має такий вигляд (Рисунок 1).

Застосувавши ППП «Statistica» для статистичних даних розподілу населення Запорізької області по віковим групам за 2018 рік рівняння лінійної регресії $Y=AX+B$ має вид: $Y=1{,}601.909 \cdot X - 6.02867$. (18)

Параметри закону: $a=43.1$, $b=1{,}601.909$. Середній вік людей, що живуть $T_ж=38.6$. При цих даних $\sum_{x=0}^{100}|F^*(x)-F(x)|=5{,}3$, де $F^*(x)$ - статистична функція розподілу, а $F(x)$ - теоретична функція розподілу.

Дані параметри $a$, $b$ можна знайти також за допомогою програми Excel, застосовуючи опцію «Пошук рішення» мінімізуючи суму $\sum_{x=0}^{100}|F^*(x)-F(x)|$.

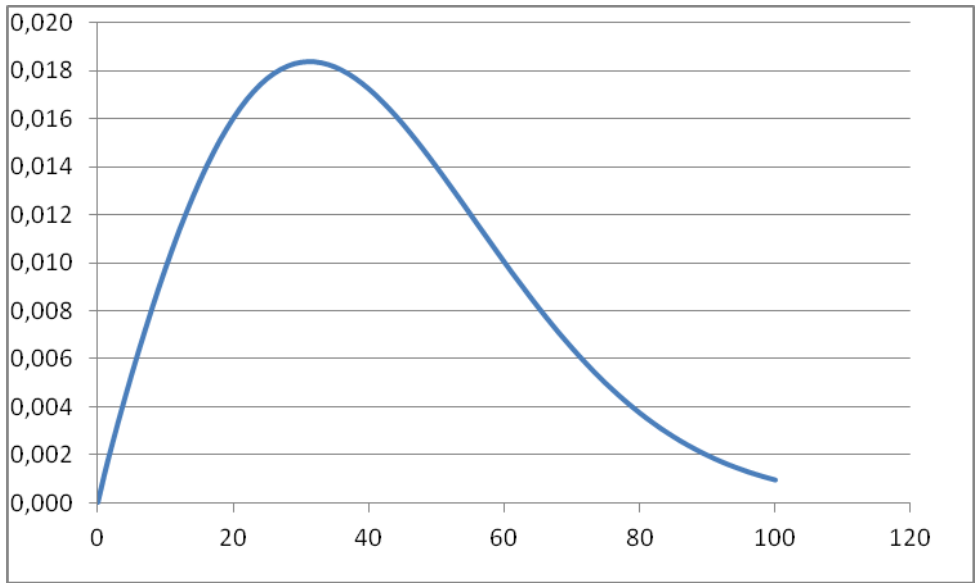

Джерело: Розробка авторів.

**Рисунок 1.** Закон Вейбулла-Гнеденко щільності розподілу ймовірності $f(t_ж)$ віку $t_ж$ людей, що живуть

Застосувавши програму Excel для статистичних даних розподілу населення Запорізької області по віковим групам за 2018 рік були зроблені такі обчислення: a=49.5, b=2.04, $T_ж$=43.8, $\sum_{x=0}^{100}|F^*(x)-F(x)| = 2.54$. Вибираємо значення параметрів $a$ і $b$ там де $\sum_{x=0}^{100}|F^*(x)-F(x)|$ менша.

$$E(T_ж) = \frac{Д_{с2}}{К_с} = \frac{60.570}{0.016263} = 3,724.291 \text{ (грн)}, \quad (18)$$

$$E_0 = \frac{E(T_ж)}{\exp\left[-\left(\frac{T_ж-c}{a}\right)^b\right]} = \frac{3,724.291}{\exp\left[-\left(\frac{43.8}{49.5}\right)^{2.04}\right]} = 8,117.411 \text{ (грн)}. \quad (19)$$

ЕЕВЖ для інших вікових груп населення Запорізької області за 2018 рік виглядають таким чином (Таблиця 4).

**Таблиця 4.** Оцінки ЕЕВЖ для Запорізької області (обидві статі, 2018 р.)

Джерело: Розробка авторів за джерелом [12].

| Вік | E(t_ж) (грн) | Вік | E(t_ж) (грн) | Вік | E(t_ж) (грн) | Вік | E(t_ж) (грн) |
|---|---|---|---|---|---|---|---|
| 0 | 8,117.411 | 26 | 6,203.638 | 51 | 2,804.481 | 76 | 737.804 |
| 1 | 8,114.577 | 27 | 6,071.571 | 52 | 2,686.592 | 77 | 691.504 |
| 2 | 8,105.764 | 28 | 5,937.295 | 53 | 2,571.426 | 78 | 647.538 |
| 3 | 8,090.801 | 29 | 5,801.078 | 54 | 2,459.061 | 79 | 605.832 |
| 4 | 8,069.620 | 30 | 5,663.183 | 55 | 2,349.563 | 80 | 566.314 |
| 5 | 8,042.196 | 31 | 5,523.877 | 56 | 2,242.990 | 81 | 528.905 |
| 6 | 8,008.537 | 32 | 5,383.420 | 57 | 2,139.389 | 82 | 493.532 |
| 7 | 7,968.676 | 33 | 5,242.073 | 58 | 2,038.797 | 83 | 460.118 |
| 8 | 7,922.667 | 34 | 5,100.089 | 59 | 1,941.242 | 84 | 428.587 |
| 9 | 7,870.586 | 35 | 4,957.721 | 60 | 1,846.744 | 85 | 398.864 |
| 10 | 7,812.526 | 36 | 4,815.214 | 61 | 1,755.314 | 86 | 370.875 |

**Таблиця 4.** (продовження)

| Вік | E(t_ж) (грн) | Вік | E(t_ж) (грн) | Вік | E(t_ж) (грн) | Вік | E(t_ж) (грн) |
|---|---|---|---|---|---|---|---|
| 11 | 7,748.598 | 37 | 4,672.806 | 62 | 1,666.955 | 87 | 344.544 |
| 12 | 7,678.931 | 38 | 4,530.731 | 63 | 1,581.660 | 88 | 319.799 |
| 13 | 7,603.666 | 39 | 4,389.213 | 64 | 1,499.419 | 89 | 296.568 |
| 14 | 7,522.962 | 40 | 4,248.470 | 65 | 1,420.211 | 90 | 274.781 |
| 15 | 7,436.990 | 41 | 4,108.712 | 66 | 1,344.010 | 91 | 254.369 |
| 16 | 7,345.934 | 42 | 3,970.137 | 67 | 1,270.785 | 92 | 235.265 |
| 17 | 7,249.990 | 43 | 3,832.938 | 68 | 1,200.496 | 93 | 217.402 |
| 18 | 7,149.366 | 44 | 3,697.295 | 69 | 1,133.100 | 94 | 200.718 |
| 19 | 7,044.278 | 45 | 3,563.381 | 70 | 1,068.551 | 95 | 185.149 |
| 20 | 6,934.953 | 46 | 3,431.355 | 71 | 1,006.794 | 96 | 170.636 |
| 21 | 6,821.627 | 47 | 3,301.370 | 72 | 947.774 | 97 | 157.121 |
| 22 | 6,704.540 | 48 | 3,173.566 | 73 | 891.429 | 98 | 144.547 |
| 23 | 6,583.942 | 49 | 3,048.072 | 74 | 837.697 | 99 | 132.862 |
| 24 | 6,460.086 | 50 | 2,925.007 | 75 | 786.512 | 100 | 122.013 |
| 25 | 6,333.231 | – | – | – | – | – | – |

По даним Таблиць 4 і 5 спостерігаємо, що результати обчислень майже співпадають. Тому запропонована методика розрахунків по формулі (16) більш проста, ніж методика розрахунків запропонованих авторами [1]. Для порівняння результатів наведемо оцінки ЕЕВЖ по аналогічним даним для деяких країн, представлені у Додатку 3 [12].

Інтегральна відсоткова ставка банків за депозитами фізичних осіб у 2018 році в Україні склала 8.59%. Оцінка ЕЕВЖ для осіб, яким виповнилось $T_ж$=42.4 років серед населення Запорізької області за 2018 рік (середньодушовими річними доходами), в розрахунках по методиці [3, 9, 11, 14] має вигляд:

$Д_{сер}$=60.570 грн,

E=ln(1+i)=ln(1+0.0859)=0.0824,

$t_0 = e_{t_ж}^0 - T_ж = 70.89 - 42.4 = 28.49$,

$$EEВЖ = 60.570 \cdot \int_0^{28,49} e^{-0.0824t} dt = 664.744 \text{ (грн).} \qquad (21)$$

На думку авторів період очікуваної тривалості майбутнього життя для осіб віку $T_ж$=42.4 років коректніше взяти з таблиці середньої очікуваної тривалості життя для осіб (обидві статі), яким виповнилося $t_ж$ років для Запорізької області (2019 р., Додаток 2).

За цією методикою маємо:

$e_{T_ж} = e_{42.4} \approx 31.32$,

$$EEВЖ = 60.570 \cdot \int_0^{31.32} e^{-0.0824t} dt = 679.357 \text{ (грн).} \qquad (22)$$

Якщо застосувати формулу (11) для довільного віку і таблицю середньої очікуваної тривалості життя для осіб (обидві статі), яким виповнилося $t_ж$ років для Запорізької області (2019 р., Додаток 2), то одержимо такі розрахунки (Таблиця 5).

**Таблиця 5.** ЕЕВЖ для Запорізької області (обидві статі, 2018 р.)

| Вік | $E(t_ж)$ (грн) | Вік | $E(t_ж)$ (грн) | Вік | $E(t_ж)$ (грн) | Вік | $E(t_ж)$ (грн) |
|---|---|---|---|---|---|---|---|
| 0 | 732.937 | 26 | 718.720 | 51 | 635.141 | 76 | 368.739 |
| 1 | 732.859 | 27 | 717.401 | 52 | 628.496 | 77 | 355.093 |
| 2 | 732.671 | 28 | 715.981 | 53 | 621.507 | 78 | 341.432 |
| 3 | 732.468 | 29 | 714.453 | 54 | 614.165 | 79 | 327.788 |
| 4 | 732.247 | 30 | 712.809 | 55 | 606.464 | 80 | 314.194 |
| 5 | 732.009 | 31 | 711.043 | 56 | 598.400 | 81 | 300.678 |
| 6 | 731.750 | 32 | 709.146 | 57 | 589.969 | 82 | 287.273 |
| 7 | 731.470 | 33 | 707.110 | 58 | 581.169 | 83 | 274.006 |
| 8 | 731.166 | 34 | 704.925 | 59 | 572.000 | 84 | 260.906 |
| 9 | 730.837 | 35 | 702.583 | 60 | 562.464 | 85 | 247.997 |
| 10 | 730.481 | 36 | 700.074 | 61 | 552.562 | 86 | 235.304 |
| 11 | 730.096 | 37 | 697.388 | 62 | 542.302 | 87 | 222.845 |
| 12 | 729.679 | 38 | 694.515 | 63 | 531.689 | 88 | 210.638 |
| 13 | 729.228 | 39 | 691.444 | 64 | 520.733 | 89 | 198.694 |
| 14 | 728.740 | 40 | 688.165 | 65 | 509.445 | 90 | 187.021 |
| 15 | 728.212 | 41 | 684.667 | 66 | 497.840 | 91 | 175.614 |
| 16 | 727.641 | 42 | 680.938 | 67 | 485.932 | 92 | 164.460 |
| 17 | 727.025 | 43 | 676.967 | 68 | 473.740 | 93 | 153.527 |
| 18 | 726.358 | 44 | 672.742 | 69 | 461.283 | 94 | 142.750 |
| 19 | 725.638 | 45 | 668.253 | 70 | 448.584 | 95 | 132.017 |
| 20 | 724.860 | 46 | 663.488 | 71 | 435.666 | 96 | 121.124 |
| 21 | 724.020 | 47 | 658.436 | 72 | 422.555 | 97 | 109.702 |
| 22 | 723.114 | 48 | 653.085 | 73 | 409.277 | 98 | 97.048 |
| 23 | 722.136 | 49 | 647.426 | 74 | 395.862 | 99 | 81.771 |
| 24 | 721.081 | 50 | 641.447 | 75 | 382.340 | 100 | 60.942 |
| 25 | 719.945 | – | – | – | – | – | – |

Якщо застосувати формулу $E(t_ж) = \dfrac{e_{t_ж}}{e_{T_ж}} \cdot \text{ЕЕВЖ}(T_ж)$ то розрахунки мають такий вигляд (Таблиця 6).

**Таблиця 6.** ЕЕВЖ для Запорізької області (обидві статі, 2018 р.)

| Вік | $E(t_ж)$ (грн) | Вік | $E(t_ж)$ (грн) | Вік | $E(t_ж)$ (грн) | Вік | $E(t_ж)$ (грн) |
|---|---|---|---|---|---|---|---|
| 0 | 1,537.611 | 26 | 1,001.770 | 51 | 525.286 | 76 | 183.326 |
| 1 | 1,528.158 | 27 | 981.347 | 52 | 508.341 | 77 | 173.698 |
| 2 | 1,506.739 | 28 | 961.001 | 53 | 491.620 | 78 | 164.400 |
| 3 | 1,485.337 | 29 | 940.734 | 54 | 475.129 | 79 | 155.431 |
| 4 | 1,463.955 | 30 | 920.551 | 55 | 458.876 | 80 | 146.788 |
| 5 | 1,442.593 | 31 | 900.456 | 56 | 442.867 | 81 | 138.468 |
| 6 | 1,421.253 | 32 | 880.454 | 57 | 427.110 | 82 | 130.467 |
| 7 | 1,399.936 | 33 | 860.550 | 58 | 411.611 | 83 | 122.782 |
| 8 | 1,378.644 | 34 | 840.747 | 59 | 396.378 | 84 | 115.406 |
| 9 | 1,357.377 | 35 | 821.052 | 60 | 381.417 | 85 | 108.336 |
| 10 | 1,336.138 | 36 | 801.470 | 61 | 366.734 | 86 | 101.563 |

**Таблиця 6.** (продовження)

| Вік | $E(t_ж)$ (грн) | Вік | $E(t_ж)$ (грн) | Вік | $E(t_ж)$ (грн) | Вік | $E(t_ж)$ (грн) |
|---|---|---|---|---|---|---|---|
| 11 | 1,314.929 | 37 | 782.006 | 62 | 352.336 | 87 | 95.082 |
| 12 | 1,293.750 | 38 | 762.665 | 63 | 338.228 | 88 | 88.882 |
| 13 | 1,272.605 | 39 | 743.453 | 64 | 324.417 | 89 | 82.954 |
| 14 | 1,251.494 | 40 | 724.377 | 65 | 310.907 | 90 | 77.287 |
| 15 | 1,230.421 | 41 | 705.442 | 66 | 297.703 | 91 | 71.864 |
| 16 | 1,209.386 | 42 | 686.654 | 67 | 284.811 | 92 | 66.668 |
| 17 | 1,188.393 | 43 | 668.020 | 68 | 272.234 | 93 | 61.671 |
| 18 | 1,167.445 | 44 | 649.546 | 69 | 259.977 | 94 | 56.838 |
| 19 | 1,146.542 | 45 | 631.239 | 70 | 248.041 | 95 | 52.110 |
| 20 | 1,125.689 | 46 | 613.105 | 71 | 236.432 | 96 | 47.398 |
| 21 | 1,104.888 | 47 | 595.152 | 72 | 225.150 | 97 | 42.546 |
| 22 | 1,084.142 | 48 | 577.387 | 73 | 214.197 | 98 | 37.272 |
| 23 | 1,063.453 | 49 | 559.816 | 74 | 203.575 | 99 | 31.044 |
| 24 | 1,042.826 | 50 | 542.447 | 75 | 193.285 | 100 | 22.782 |
| 25 | 1,022.264 | – | – | – | – | – | – |

Валовий регіональний продукт на душу населення в Запорізькій області за 2018 рік (попередні дані) складає приблизно 91 тис грн. Оцінка ЕЕВЖ для осіб яким виповнилось $T_ж$=42.4 років населення Запорізької області за 2018 рік (валовий регіональний продукт на душу населення) в розрахунках має вигляд:

$$EEBЖ = 91.000 \cdot \int_0^{31.32} e^{-0.0824 t} dt = 1,020.661 \text{ (грн)}. \qquad (24)$$

Якщо застосувати формулу (11) для довільного віку і таблицю середньої очікуваної тривалості життя для осіб (обидві статі), яким виповнилося tж років для Запорізької області (2019 р., Додаток 2), то одержимо такі розрахунки (Таблиця 7).

**Таблиця 7.** ЕЕВЖ для Запорізької області (обидві статі, 2018 р.)

Джерело: Розробка авторів за джерелом [12].

| Вік | $E(t_ж)$ (грн) | Вік | $E(t_ж)$ (грн) | Вік | $E(t_ж)$ (грн) | Вік | $E(t_ж)$ (грн) |
|---|---|---|---|---|---|---|---|
| 0 | 1,101.160 | 26 | 1,079.800 | 51 | 954.231 | 76 | 553.992 |
| 1 | 1,101.043 | 27 | 1,077.818 | 52 | 944.249 | 77 | 533.490 |
| 2 | 1,100.761 | 28 | 1,075.685 | 53 | 933.748 | 78 | 512.966 |
| 3 | 1,100.455 | 29 | 1,073.389 | 54 | 922.717 | 79 | 492.467 |
| 4 | 1,100.124 | 30 | 1,070.921 | 55 | 911.148 | 80 | 472.043 |
| 5 | 1,099.765 | 31 | 1,068.267 | 56 | 899.033 | 81 | 451.737 |
| 6 | 1,099.377 | 32 | 1,065.417 | 57 | 886.366 | 82 | 431.597 |
| 7 | 1,098.955 | 33 | 1,062.358 | 58 | 873.145 | 83 | 411.666 |
| 8 | 1,098.499 | 34 | 1,059.075 | 59 | 859.370 | 84 | 391.984 |
| 9 | 1,098.005 | 35 | 1,055.557 | 60 | 845.042 | 85 | 372.590 |
| 10 | 1,097.471 | 36 | 1,051.787 | 61 | 830.167 | 86 | 353.519 |
| 11 | 1,096.892 | 37 | 1,047.752 | 62 | 814.751 | 87 | 334.801 |
| 12 | 1,096.266 | 38 | 1,043.435 | 63 | 798.806 | 88 | 316.461 |
| 13 | 1,095.588 | 39 | 1,038.822 | 64 | 782.346 | 89 | 298.517 |

**Таблиця 7.** (продовження)

| Вік | $E(t_ж)$ (грн) | Вік | $E(t_ж)$ (грн) | Вік | $E(t_ж)$ (грн) | Вік | $E(t_ж)$ (грн) |
|---|---|---|---|---|---|---|---|
| 14 | 1,094.855 | 40 | 1,033.895 | 65 | 765.388 | 90 | 280.979 |
| 15 | 1,094.062 | 41 | 1,028.639 | 66 | 747.952 | 91 | 263.842 |
| 16 | 1,093.204 | 42 | 1,023.037 | 67 | 730.061 | 92 | 247.084 |
| 17 | 1,092.277 | 43 | 1,017.071 | 68 | 711.744 | 93 | 230.657 |
| 18 | 1,091.276 | 44 | 1.010724 | 69 | 693.029 | 94 | 214.466 |
| 19 | 1,090.194 | 45 | 1,003.980 | 70 | 673.950 | 95 | 198.341 |
| 20 | 1,089.025 | 46 | 996.820 | 71 | 654.542 | 96 | 181.976 |
| 21 | 1,087.763 | 47 | 989.230 | 72 | 634.843 | 97 | 164.815 |
| 22 | 1,086.402 | 48 | 981.191 | 73 | 614.896 | 98 | 145.804 |
| 23 | 1,084.933 | 49 | 972.688 | 74 | 594.741 | 99 | 122.853 |
| 24 | 1,083.348 | 50 | 963.707 | 75 | 574.425 | 100 | 91.559 |
| 25 | 1,081.641 | – | – | – | – | – | – |

Якщо застосувати формулу $E(t_ж) = \dfrac{e_{t_ж}}{e_{T_ж}} \cdot \text{ЕЕВЖ}(Т_ж)$, то розрахунки мають такий вигляд (Таблиця 8).

**Таблиця 8.** ЕЕВЖ для Запорізької області (обидві статі, 2018 р.)

*Джерело: Розробка авторів за джерелом [12].*

| Вік | $E(t_ж)$ (грн) | Вік | $E(t_ж)$ (грн) | Вік | $E(t_ж)$ (грн) | Вік | $E(t_ж)$ (грн) |
|---|---|---|---|---|---|---|---|
| 0 | 2,310.095 | 26 | 1,505.052 | 51 | 789.186 | 76 | 275.427 |
| 1 | 2,295.894 | 27 | 1,474.369 | 52 | 763.729 | 77 | 260.963 |
| 2 | 2,263.713 | 28 | 1,443.801 | 53 | 738.606 | 78 | 246.994 |
| 3 | 2,231.560 | 29 | 1,413.352 | 54 | 713.830 | 79 | 233.519 |
| 4 | 2,199.435 | 30 | 1,383.029 | 55 | 689.411 | 80 | 220.533 |
| 5 | 2,167.341 | 31 | 1,352.839 | 56 | 665.360 | 81 | 208.033 |
| 6 | 2,135.280 | 32 | 1,322.788 | 57 | 641.686 | 82 | 196.013 |
| 7 | 2,103.254 | 33 | 1,292.883 | 58 | 618.401 | 83 | 184.466 |
| 8 | 2,071.264 | 34 | 1,263.133 | 59 | 595.515 | 84 | 173.386 |
| 9 | 2,039.313 | 35 | 1.233.543 | 60 | 573.038 | 85 | 162.763 |
| 10 | 2,007.404 | 36 | 1,204.123 | 61 | 550.978 | 86 | 152.588 |
| 11 | 1,975.539 | 37 | 1,174.880 | 62 | 529.346 | 87 | 142.850 |
| 12 | 1,943.721 | 38 | 1,145.822 | 63 | 508.151 | 88 | 133.536 |
| 13 | 1,911.952 | 39 | 1,116.959 | 64 | 487.401 | 89 | 124.630 |
| 14 | 1,880.236 | 40 | 1,088.298 | 65 | 467.104 | 90 | 116.115 |
| 15 | 1,848.575 | 41 | 1,059.850 | 66 | 447.267 | 91 | 107.968 |
| 16 | 1,816.973 | 42 | 1,031.623 | 67 | 427.898 | 92 | 100.161 |
| 17 | 1,785.434 | 43 | 1,003.628 | 68 | 409.003 | 93 | 92.654 |
| 18 | 1,753.960 | 44 | 975.873 | 69 | 390.587 | 94 | 85.393 |
| 19 | 1,722.557 | 45 | 948.368 | 70 | 372.656 | 95 | 78.290 |
| 20 | 1,691.227 | 46 | 921.125 | 71 | 355.213 | 96 | 71.211 |
| 21 | 1,659.975 | 47 | 894.152 | 72 | 338.263 | 97 | 63.920 |
| 22 | 1,628.806 | 48 | 867.462 | 73 | 321.808 | 98 | 55.998 |
| 23 | 1,597.724 | 49 | 841.063 | 74 | 305.850 | 99 | 46.640 |
| 24 | 1,566.735 | 50 | 814.968 | 75 | 290.389 | 100 | 34.228 |
| 25 | 1,535.842 | – | – | – | – | – | – |

# ВИСНОВКИ

На теперішній час в Україні досі немає унормованого методологічного підходу до оцінки вартості статистичного життя, який би сприяв розрахункам соціальних виплат у разі нещасних випадків. При розрахунку цього показника важливо визначити інститути відповідальності, якими на думку авторів доцільно було б призначити Міністерство соціальної політики України, Міністерство економічного розвитку і торгівлі України, а також Інститут демографії та соціальних досліджень імені Птухи НАН України.

В Європі рекомендоване значення вартості життя людини знаходиться в діапазоні від $ 0.9 млн до $ 16.2 млн з середнім значенням $ 5.4 млн [2]. Розрахунки зарубіжних фахівців доводять, що чим повніше оцінюється вартість людського життя тим більш значущу вигоду має спільнота. За оцінками шведських вчених, якщо економічний еквівалент врятованого життя становить в середньому 1 млн євро, то при умові збереження 45 тис життів протягом шести років ефект дорівнює приблизно 22.5 млрд євро, при чому демографічна складова цього ефекту, згідно з розрахунками, дасть приріст населення до сотень тисяч людей за 10 років [2].

Аналіз ЕЕВЖ західних країн показує, що відношення ЕЕВЖ до середньодушового наявного грошового річного доходу знаходиться середньо в діапазоні пропорцій 50:1-100:1. Тому в Україні для визначення державних або корпоративних виплат сім'ям загиблих при надзвичайних ситуаціях у страхових сумах системи особистого страхування та страхування відповідальності рекомендується використовувати значення ЕЕВЖ згідно даного діапазону, тобто від 2.5 млн грн до 5 млн грн.

Ключовий висновок дослідження базується на твердженні про те, що статистична вартість життя людини в Україні недооцінена у теперішній час. Крім трагічних подій у людському житті, це призводить до істотних наслідків, які впливають на макроекономічні показники у національній економіці: людський капітал знижується, збитки від нещасних випадків та захворювань не компенсуються у достатній мірі, ризики втрат від дорожньо-транспортних пригод підвищуються, тощо. ЕЕВЖ має використовуватись також для удосконалення заходів з підвищення безпеки; для покращення функціонування правоохоронної системи, системи охорони здоров'я, попередження надзвичайних ситуацій; для поліпшення розрахунків у страховому бізнесі. Реальні оцінки ЕЕВЖ та впровадження їх у національне законодавство має стати не тільки актуальним науковим завданням, але і важливою практичною задачею.

Подальші дослідження варто спрямувати на встановлення кількісного взаємозв'язку між показником ЕЕВЖ та розвитком трудового потенціалу в Україні та обґрунтування способів підвищення цього потенціалу на основі теорії людського капіталу з використанням розрахунків, які відштовхувалися б від рівня річної заробітної платні середньостатистичної людини.

# СПИСОК ЛІТЕРАТУРИ

1. Aivar, L. K., Trunov, Y. L., & Kharysov, H. Kh. (2006). Ekvyvalent stoymosty chelovecheskoi zhyzny [Equivalent to the cost of human life]. *Predstavytelnaia vlast: zakonodatelstvo, kommentaryy, problemy - Representative power: legislation, comments, problems, 3*, 24-29. (In Russian). Retrieved from http://www.pvlast.ru/img/pdf2006-3/8.pdf
2. Berlin, M., Hladun, A., Lysenko, Yu., & Shchetinina E. (2016). Bestsennye liudy:zachem znat stoymost chelovecheskoi zhyzny [Priceless people: why know the value of human life]. *Forbes Ukraine*.
3. Bykov, A. (2007). Pro metodolohii otsinky vartosti zhyttia serednostatystychnoi liudyny [On the methodology of estimating the cost of living of the average person]. *Strakhova sprava, 3*, 10-25. (In Ukrainian)
4. Carlson, I. W. (1963). *Valuation of Life Saving* (Ph.D. Thesis). Harward University. Cambridge.
5. Derzhavna sluzhba statystyky Ukrainy (2013). *Metodolohichni polozhennia zi statystychnoho analizu pryrodnoho rukhu naselennia [Methodological provisions for statistical analysis of the natural movement of the population]*. (In Ukrainian). Retrieved from http://database.ukrcensus.gov.ua/PXWEB2007/ukr/method/%D0%9F%D1%80%D0%B8%D1%80%D0%BE%D0%B4_%D1%80%D1%83 %D1%85_08-02-13.pdf

6. Kabinet Ministriv Ukrainy (1996). *Pro zatverdzhennia Polozhennia pro oboviazkove osobyste strakhuvannia vid neshchasnykh vypadkiv na transporti [On approval of the Regulations on compulsory personal accident insurance on transport]*. (In Ukrainian). Retrieved from http://zakon2.rada.gov.ua/laws/show/959-96-п
7. Kabinet Ministriv Ukrainy (2007). *Pro zatverdzhennia Poriadku ta umov vyplaty odnorazovoi hroshovoi dopomohy u razi zahybeli (smerti), poranennia (kontuzii, travmy abo kalitstva) chy invalidnosti spivrobitnykiv kadrovoho skladu rozviduvalnykh orhaniv [About Approval of the Procedure and Conditions for Payment of Lump sum Financial Assistance in the Case of Death, Injury or Disability of Intelligence Staff]*. (In Ukrainian). Retrieved from https://zakon.rada.gov.ua/laws/show/1331-2007-%D0%BF
8. Kabinet Ministriv Ukrainy. (2013) *Pro zatverdzhennia Poriadku pryznachennia i vyplaty odnorazovoi hroshovoi dopomohy u razi zahybeli (smerti), invalidnosti abo chastkovoi vtraty pratsezdatnosti bez vstanovlennia invalidnosti viiskovosluzhbovtsiv, viiskovozo-zaboviazanykh ta rezervistiv, yaki pryzvani na navchalni (abo perevirochni) ta spetsialni zbory chy dlia prokhodzhennia sluzhby u viiskovomu rezervi [On approval of the Procedure for the appointment and payment of one-off financial assistance in the event of death (death), disability or partial disability without establishing the disability of servicemen, servicemen and reservists who are called for training (or checking) and special training or military training or special training]*. (In Ukrainian). Retrieved from http://zakon3.rada.gov.ua/laws/show/975-2013-%D0%BF
9. Kaneva, T. V., & Kartashova, S. S. (2014). Ekonomichnyi ekvivalent otsinky vartosti serednostatystychnoho zhyttia v Ukraini: metodolohiia, rekomendatsii [The economic equivalent of estimating the value of the average life in Ukraine: methodology, recommendations]. *Statystyka Ukrainy - Statistics of Ukraine, 3*(66), 31-37. (In Ukrainian)
10. Khan, H., & Shapyro, S. (1969). *Statisticheskie modeli v inzhenernykh zadachakh [Statistical models in engineering problems]*. Moskva: Mir. (In Russian)
11. Levytskyi, S. I., Hnieushev, O. M., & Makhlynets, V. M. (2018). *Aktuarnyi pidkhid do modeliuvannia ekonomichnoho ekvivalentu vartosti zhyttia u Zaporizkyi oblasti [Actuarial approach to modeling the economic equivalent of cost of living in Zaporizhzhia region]*. Kyiv: KNEU. (In Ukrainian). Retrieved from http://ir.kneu.edu.ua/bitstream/handle/2010/25982/ZE_2018_68.pdf
12. Levytskyi, S. I., Hnieushev, O. M., & Makhlynets, V. M. (2018). Modeling of the economic equivalent of the cost of living in the Zaporizhzhia region. *Skhidna Yevropa: ekonomika, biznes ta upravlinnia*, 6(17), 813-818. (In Ukrainian). Retrieved from http://www.easterneu-rope-ebm.in.ua/journal/17_2018/141.pdf
13. Rynhach, N. A. (2016). Economic equivalent of losses due to of premature mortality in Ukraine. *Demography and Social Economy, 2*, 39-49. (In Ukrainian). Retrieved from http://nbuv.gov.ua/UJRN/dse_2016_2_5
14. Shevchuk, O. (2014). Methodological approaches for assessing the economic equivalent of the human life value in Ukraine. *Rehionalna ekonomika - Regional economy, 2*, 74-83. (In Ukrainian). Retrieved from http://nbuv.gov.ua/UJRN/regek_2014_2_10
15. Tretiakova, H. (2013). Zrostannia «vartosti zhyttia» yak faktor (stymul) zrostannia strakhuvannia vidpovidalnosti v Ukraini [The rise in the cost of living as a factor (stimulus) for the growth of liability insurance in Ukraine]. In *XIII Mezhdunarodnyy Yaltinskiy finansovyy forum - XIII International Yalta Financial Forum*. (In Russian). Retrieved from http://ufu.org.ua/files/dagest/20-09-13.ppt

**Додаток 1**

Модельні розрахунки коефіцієнту дожиття в Запорізькій області, 2018 рік

| Вік (років) | Число осіб, які доживають до віку X років | Число осіб, які вмирають у віці від X до X+1 | Ймовірність померти у віці від X до X+1 | Ймовірність дожити у віці від X до X+1 | Число осіб, які живуть у віці від X до X+1 | Число чоловіко-років від X років і старше | Середня очікувана тривалість життя | Коефіцієнт дожиття |
|---|---|---|---|---|---|---|---|---|
| x | $l_x$ | $d_x$ | $q_x$ | $p_x$ | $L_x$ | $T_x$ | $e_x^0$ | $P_x$ |
| 0 | 100,000 | 795 | 0.00795 | 0.99205 | 99.602 | 7,088.757 | 70.89 | 0.99592 |
| 1 | 99.205 | 18 | 0.00018 | 0.99982 | 99.196 | 6,989.155 | 70.45 | 0.99981 |
| 2 | 99.187 | 19 | 0.00019 | 0.99981 | 99.177 | 6,889.959 | 69.46 | 0.99980 |
| 3 | 99.168 | 21 | 0.00021 | 0.99979 | 99.157 | 6,790.781 | 68.48 | 0.99978 |
| 4 | 99.147 | 22 | 0.00023 | 0.99977 | 99.136 | 6,691.624 | 67.49 | 0.99976 |
| 5 | 99.125 | 24 | 0.00024 | 0.99976 | 99.113 | 6,592.488 | 66.51 | 0.99975 |
| 6 | 99.100 | 26 | 0.00026 | 0.99974 | 99.087 | 6,493.375 | 65.52 | 0.99972 |
| 7 | 99.074 | 28 | 0.00029 | 0.99971 | 99.060 | 6,394.288 | 64.54 | 0.99970 |
| 8 | 99.046 | 31 | 0.00031 | 0.99969 | 99.030 | 6,295.228 | 63.56 | 0.99968 |
| 9 | 99.015 | 33 | 0.00034 | 0.99966 | 98.998 | 6,196.198 | 62.58 | 0.99965 |
| 10 | 98.982 | 36 | 0.00036 | 0.99964 | 98.964 | 6,097.199 | 61.60 | 0.99962 |
| 11 | 98.946 | 39 | 0.00039 | 0.99961 | 98.926 | 5,998.236 | 60.62 | 0.99959 |
| 12 | 98.907 | 42 | 0.00042 | 0.99958 | 98.886 | 5,899.309 | 59.65 | 0.99956 |
| 13 | 98.865 | 45 | 0.00046 | 0.99954 | 98.842 | 5,800.423 | 58.67 | 0.99952 |
| 14 | 98.820 | 49 | 0.00050 | 0.99950 | 98.795 | 5,701.581 | 57.70 | 0.99948 |
| 15 | 98.770 | 53 | 0.00054 | 0.99946 | 98.744 | 5,602.786 | 56.73 | 0.99944 |
| 16 | 98.717 | 57 | 0.00058 | 0.99942 | 98.689 | 5,504.042 | 55.76 | 0.99939 |
| 17 | 98.660 | 62 | 0.00063 | 0.99937 | 98.629 | 5,405.354 | 54.79 | 0.99934 |
| 18 | 98.598 | 67 | 0.00068 | 0.99932 | 98.564 | 5,306.725 | 53.82 | 0.99929 |
| 19 | 98.530 | 73 | 0.00074 | 0.99926 | 98.494 | 5,208.161 | 52.86 | 0.99923 |
| 20 | 98.458 | 79 | 0.00080 | 0.99920 | 98.419 | 5,109.667 | 51.90 | 0.99917 |
| 21 | 98.379 | 85 | 0.00086 | 0.99914 | 98.337 | 5,011.248 | 50.94 | 0.99910 |
| 22 | 98.295 | 92 | 0.00093 | 0.99907 | 98.249 | 4,912.911 | 49.98 | 0.99903 |
| 23 | 98.203 | 99 | 0.00101 | 0.99899 | 98.153 | 4,814.662 | 49.03 | 0.99895 |
| 24 | 98.104 | 107 | 0.00109 | 0.99891 | 98.050 | 4,716.509 | 48.08 | 0.99886 |
| 25 | 97.996 | 116 | 0.00118 | 0.99882 | 97.939 | 4,618.459 | 47.13 | 0.99877 |
| 26 | 97.881 | 125 | 0.00128 | 0.99872 | 97.818 | 4,520.521 | 46.18 | 0.99867 |
| 27 | 97.756 | 135 | 0.00138 | 0.99862 | 97.688 | 4,422.703 | 45.24 | 0.99856 |
| 28 | 97.620 | 146 | 0.00150 | 0.99850 | 97.547 | 4,325.015 | 44.30 | 0.99844 |
| 29 | 97.474 | 158 | 0.00162 | 0.99838 | 97.395 | 4,227.467 | 43.37 | 0.99831 |
| 30 | 97.316 | 170 | 0.00175 | 0.99825 | 97.231 | 4,130.072 | 42.44 | 0.99818 |
| 31 | 97.146 | 184 | 0.00189 | 0.99811 | 97.054 | 4,032.841 | 41.51 | 0.99803 |
| 32 | 96.962 | 199 | 0.00205 | 0.99795 | 96.863 | 3,935.787 | 40.59 | 0.99787 |
| 33 | 96.763 | 215 | 0.00222 | 0.99778 | 96.656 | 3,838.924 | 39.67 | 0.99769 |
| 34 | 96.549 | 232 | 0.00240 | 0.99760 | 96.433 | 3,742.269 | 38.76 | 0.99750 |
| 35 | 96.317 | 250 | 0.00260 | 0.99740 | 96.192 | 3,645.836 | 37.85 | 0.99730 |
| 36 | 96.067 | 270 | 0.00281 | 0.99719 | 95.932 | 3,549.644 | 36.95 | 0.99708 |
| 37 | 95.797 | 291 | 0.00304 | 0.99696 | 95.652 | 3,453.712 | 36.05 | 0.99684 |
| 38 | 95.506 | 314 | 0.00329 | 0.99671 | 95.349 | 3,358.060 | 35.16 | 0.99658 |
| 39 | 95.192 | 338 | 0.00356 | 0.99644 | 95.023 | 3,262.711 | 34.27 | 0.99630 |
| 40 | 94.854 | 365 | 0.00385 | 0.99615 | 94.671 | 3,167.688 | 33.40 | 0.99600 |
| 41 | 94.489 | 393 | 0.00416 | 0.99584 | 94.292 | 3,073.017 | 32.52 | 0.99567 |
| 42 | 94.096 | 424 | 0.00450 | 0.99550 | 93.884 | 2,978.725 | 31.66 | 0.99531 |
| 43 | 93.672 | 456 | 0.00487 | 0.99513 | 93.444 | 2,884.841 | 30.80 | 0.99493 |
| 44 | 93.216 | 491 | 0.00527 | 0.99473 | 92.970 | 2,791.398 | 29.95 | 0.99451 |
| 45 | 92.724 | 529 | 0.00570 | 0.99430 | 92.460 | 2,698.428 | 29.10 | 0.99407 |
| 46 | 92.196 | 569 | 0.00617 | 0.99383 | 91.911 | 2,605.968 | 28.27 | 0.99358 |
| 47 | 91.627 | 611 | 0.00667 | 0.99333 | 91.321 | 2,514.056 | 27.44 | 0.99305 |
| 48 | 91.015 | 657 | 0.00722 | 0.99278 | 90.687 | 2,422.735 | 26.62 | 0.99249 |
| 49 | 90.358 | 706 | 0.00781 | 0.99219 | 90.006 | 2,332.048 | 25.81 | 0.99187 |
| 50 | 89.653 | 758 | 0.00845 | 0.99155 | 89.274 | 2,242.043 | 25.01 | 0.99121 |

**Додаток 1.** (продовження)

| Вік (років) | Число осіб, які доживають до віку X років | Число осіб, які вмирають у віці від X до X+1 | Ймовірність померти у віці від X до X+1 | Ймовірність дожити у віці від X до X+1 | Число осіб, які живуть у віці від X до X+1 | Число чоловіко-років від X років і старше | Середня очікувана тривалість життя | Коефіцієнт дожиття |
|---|---|---|---|---|---|---|---|---|
| 51 | 88.895 | 813 | 0.00914 | 0.99086 | 88.489 | 2,152.769 | 24.22 | 0.99049 |
| 52 | 88.082 | 871 | 0.00989 | 0.99011 | 87.647 | 2,064.280 | 23.44 | 0.98971 |
| 53 | 87.211 | 933 | 0.01070 | 0.98930 | 86.745 | 1,976.633 | 22.66 | 0.98886 |
| 54 | 86.278 | 999 | 0.01158 | 0.98842 | 85.779 | 1,889.888 | 21.90 | 0.98795 |
| 55 | 85.279 | 1.068 | 0.01252 | 0.98748 | 84.745 | 1,804.110 | 21.16 | 0.98697 |
| 56 | 84.211 | 1.141 | 0.01355 | 0.98645 | 83.641 | 1,719.364 | 20.42 | 0.98590 |
| 57 | 83.070 | 1.218 | 0.01466 | 0.98534 | 82.462 | 1,635.723 | 19.69 | 0.98475 |
| 58 | 81.853 | 1.298 | 0.01586 | 0.98414 | 81.204 | 1,553.262 | 18.98 | 0.98350 |
| 59 | 80.555 | 1.382 | 0.01716 | 0.98284 | 79.864 | 1,472.058 | 18.27 | 0.98215 |
| 60 | 79.173 | 1.469 | 0.01856 | 0.98144 | 78.438 | 1,392.194 | 17.58 | 0.98069 |
| 61 | 77.703 | 1.560 | 0.02008 | 0.97992 | 76.923 | 1,313.756 | 16.91 | 0.97911 |
| 62 | 76.143 | 1.654 | 0.02172 | 0.97828 | 75.316 | 1,236.832 | 16.24 | 0.97740 |
| 63 | 74.489 | 1.751 | 0.02350 | 0.97650 | 73.614 | 1,161.516 | 15.59 | 0.97555 |
| 64 | 72.738 | 1.849 | 0.02543 | 0.97457 | 71.814 | 1,087.903 | 14.96 | 0.97355 |
| 65 | 70.889 | 1.950 | 0.02751 | 0.97249 | 69.914 | 1,016.089 | 14.33 | 0.97138 |
| 66 | 68.939 | 2.052 | 0.02976 | 0.97024 | 67.913 | 946.175 | 13.72 | 0.96904 |
| 67 | 66.887 | 2.154 | 0.03220 | 0.96780 | 65.811 | 878.262 | 13.13 | 0.96651 |
| 68 | 64.734 | 2.255 | 0.03483 | 0.96517 | 63.606 | 812.452 | 12.55 | 0.96377 |
| 69 | 62.479 | 2.354 | 0.03768 | 0.96232 | 61.302 | 748.845 | 11.99 | 0.96080 |
| 70 | 60.125 | 2.451 | 0.04077 | 0.95923 | 58.899 | 687.543 | 11.44 | 0.95760 |
| 71 | 57.673 | 2.544 | 0.04411 | 0.95589 | 56.402 | 628.644 | 10.90 | 0.95413 |
| 72 | 55.130 | 2.631 | 0.04772 | 0.95228 | 53.814 | 572.243 | 10.38 | 0.95038 |
| 73 | 52.499 | 2.710 | 0.05162 | 0.94838 | 51.144 | 518.428 | 9.87 | 0.94632 |
| 74 | 49.789 | 2.781 | 0.05585 | 0.94415 | 48.399 | 467.284 | 9.39 | 0.94193 |
| 75 | 47.008 | 2.840 | 0.06042 | 0.93958 | 45.588 | 418.886 | 8.91 | 0.93718 |
| 76 | 44.168 | 2.887 | 0.06537 | 0.93463 | 42.724 | 373.298 | 8.45 | 0.93205 |
| 77 | 41.281 | 2.919 | 0.07072 | 0.92928 | 39.821 | 330.573 | 8.01 | 0.92649 |
| 78 | 38.362 | 2.935 | 0.07651 | 0.92349 | 36.894 | 290.752 | 7.58 | 0.92048 |
| 79 | 35.427 | 2.932 | 0.08277 | 0.91723 | 33.960 | 253.858 | 7.17 | 0.91399 |
| 80 | 32.494 | 2.910 | 0.08955 | 0.91045 | 31.039 | 219.898 | 6.77 | 0.90696 |
| 81 | 29.584 | 2.866 | 0.09688 | 0.90312 | 28.151 | 188.858 | 6.38 | 0.89936 |
| 82 | 26.718 | 2.800 | 0.10481 | 0.89519 | 25.318 | 160.707 | 6.01 | 0.89114 |
| 83 | 23.918 | 2.712 | 0.11339 | 0.88661 | 22.562 | 135.389 | 5.66 | 0.88225 |
| 84 | 21.206 | 2.601 | 0.12267 | 0.87733 | 19.905 | 112.827 | 5.32 | 0.87263 |
| 85 | 18.605 | 2.469 | 0.13272 | 0.86728 | 17.370 | 92.921 | 4.99 | 0.86224 |
| 86 | 16.135 | 2.317 | 0.14358 | 0.85642 | 14.977 | 75.551 | 4.68 | 0.85100 |
| 87 | 13.819 | 2.147 | 0.15533 | 0.84467 | 12.745 | 60.574 | 4.38 | 0.83884 |
| 88 | 11.672 | 1.962 | 0.16805 | 0.83195 | 10.691 | 47.829 | 4.10 | 0.82570 |
| 89 | 9.711 | 1.765 | 0.18181 | 0.81819 | 8.828 | 37.137 | 3.82 | 0.81149 |
| 90 | 7.945 | 1.563 | 0.19669 | 0.80331 | 7.164 | 28.309 | 3.56 | 0.79613 |
| 91 | 6.382 | 1.358 | 0.21279 | 0.78721 | 5.703 | 21.146 | 3.31 | 0.77953 |
| 92 | 5.024 | 1.157 | 0.23022 | 0.76978 | 4.446 | 15.442 | 3.07 | 0.76159 |
| 93 | 3.868 | 963 | 0.24906 | 0.75094 | 3.386 | 10.996 | 2.84 | 0.74219 |
| 94 | 2.904 | 783 | 0.26945 | 0.73055 | 2.513 | 7.610 | 2.62 | 0.72124 |
| 95 | 2.122 | 619 | 0.29151 | 0.70849 | 1.812 | 5.097 | 2.40 | 0.69859 |
| 96 | 1.503 | 474 | 0.31538 | 0.68462 | 1.266 | 3.285 | 2.19 | 0.67413 |
| 97 | 1.029 | 351 | 0.34119 | 0.65881 | 854 | 2.019 | 1.96 | 0.64771 |
| 98 | 678 | 250 | 0.36913 | 0.63087 | 553 | 1.165 | 1.72 | 0.61918 |
| 99 | 428 | 171 | 0.39934 | 0.60066 | 342 | 612 | 1.43 | 0.78827 |
| 100 | 257 | 111 | 0.43204 | 0.56796 | 270 | 270 | 1.05 | 1.00353 |
| – | 146 | – | – | – | – | – | – | – |

**Додаток 2**

Середня очікувана тривалість життя в Запорізькій області, 2018 рік. Міські поселення та сільська місцевість, обидві статі

| Вік (років) | Середня очікувана тривалість життя | Вік (років) | Середня очікувана тривалість життя | Вік (років) | Середня очікувана тривалість життя | Вік (років) | Середня очікувана тривалість життя |
|---|---|---|---|---|---|---|---|
| $x$ | $e_x^0$ | $x$ | $e_x^0$ | $x$ | $e_x^0$ | $x$ | $e_x^0$ |
| 0 | 70.89 | 26 | 46.18 | 51 | 24.22 | 76 | 8.45 |
| 1 | 70.45 | 27 | 45.24 | 52 | 23.44 | 77 | 8.01 |
| 2 | 69.46 | 28 | 44.30 | 53 | 22.66 | 78 | 7.58 |
| 3 | 68.48 | 29 | 43.37 | 54 | 21.90 | 79 | 7.17 |
| 4 | 67.49 | 30 | 42.44 | 55 | 21.16 | 80 | 6.77 |
| 5 | 66.51 | 31 | 41.51 | 56 | 20.42 | 81 | 6.38 |
| 6 | 65.52 | 32 | 40.59 | 57 | 19.69 | 82 | 6.01 |
| 7 | 64.54 | 33 | 39.67 | 58 | 18.98 | 83 | 5.66 |
| 8 | 63.56 | 34 | 38.76 | 59 | 18.27 | 84 | 5.32 |
| 9 | 62.58 | 35 | 37.85 | 60 | 17.58 | 85 | 4.99 |
| 10 | 61.60 | 36 | 36.95 | 61 | 16.91 | 86 | 4.68 |
| 11 | 60.62 | 37 | 36.05 | 62 | 16.24 | 87 | 4.38 |
| 12 | 59.65 | 38 | 35.16 | 63 | 15.59 | 88 | 4.10 |
| 13 | 58.67 | 39 | 34.27 | 64 | 14.96 | 89 | 3.82 |
| 14 | 57.70 | 40 | 33.40 | 65 | 14.33 | 90 | 3.56 |
| 15 | 56.73 | 41 | 32.52 | 66 | 13.72 | 91 | 3.31 |
| 16 | 55.76 | 42 | 31.66 | 67 | 13.13 | 92 | 3.07 |
| 17 | 54.79 | 43 | 30.80 | 68 | 12.55 | 93 | 2.84 |
| 18 | 53.82 | 44 | 29.95 | 69 | 11.99 | 94 | 2.62 |
| 19 | 52.86 | 45 | 29.10 | 70 | 11.44 | 95 | 2.40 |
| 20 | 51.90 | 46 | 28.27 | 71 | 10.90 | 96 | 2.19 |
| 21 | 50.94 | 47 | 27.44 | 72 | 10.38 | 97 | 1.96 |
| 22 | 49.98 | 48 | 26.62 | 73 | 9.87 | 98 | 1.72 |
| 23 | 49.03 | 49 | 25.81 | 74 | 9.39 | 99 | 1.43 |
| 24 | 48.08 | 50 | 25.01 | 75 | 8.91 | 100 | 1.05 |
| 25 | 47.13 | – | – | – | – | – | – |

Розрахунок ЕЕВЖ для різних країн світу

| Параметри | | Країна, рік | | | | | | | |
|---|---|---|---|---|---|---|---|---|---|
| Умовне позначення | Найменування | Росія, 2006 | Німеччина, 2005 | Велико-британія, 2005 | Франція, 2005 | Нідерланди, 2005 | США, 2004 | Швеція, 2005 | Португалія, 2005 |
| $Д_{с2}$ | Середньодушовий наявний грошовий річний дохід | 104.832 рублів | 21.329 євро | 24.715 євро | 22.790 євро | 24.486 євро | 34.675 дол | 23.942 євро | 13.881 євро |
| $P_y$ | Фоновий ризик смерті людей (загальний коефіцієнт смертності Кс) | 0.0152 | 0.0099 | 0.0097 | 0.0084 | 0.0084 | 0.0084 | 0.0101 | 0.0097 |
| $T_ж$ | Середній вік людей, що живуть (роки) | 38.46 | 41.2 | 36.79 | 38.89 | 38.46 | 35.83 | 40.42 | 39.6 |
| Параметри щільності розподілу ймовірностей віку живуть людей: | | | | | | | | | |
| $a$ | Параметр масштабу | 43.31 | 46.4 | 40.97 | 43.65 | 43.26 | 39.82 | 45.47 | 44.55 |
| $b$ | Параметр форми | 1.86 | 1.91 | 1.58 | 1.73 | 1.79 | 1.55 | 1.78 | 1.82 |
| $c$ | Параметр зсуву | 0 | 0 | 0 | 0 | 0 | 0 | 0 | 0 |
| $E(T_ж)$ | Економічний еквівалент життя середньостатистичної людини віку Тж (в національних грошових одиницях (млн) в євро (млн) за обмінним курсом) | 6.90 0.20 | 2.15 | 2.55 | 2.71 | 2.92 | 4.13 3.33 | 2.37 | 1.43 |
| $E(t_ж)$ | Економічний еквівалент життя середньостатистичної людини у віці $t_ж$ років (млн нац. грошових одиниць) | – | – | – | – | – | – | – | – |
| $E_0$ | 0 років | 15.38 | 4.77 | 5.93 | 6.15 | 6.56 | 9.65 | 5.33 | 3.2 |
| $E_{10}$ | 10 років | 14.4 | 4.52 | 5.32 | 5.69 | 6.1 | 8.58 | 4.98 | 3 |
| $E_{20}$ | 20 років | 12.12 | 3.9 | 4.29 | 4.75 | 5.1 | 6.84 | 4.23 | 2.54 |
| $E_{30}$ | 30 років | 9.28 | 3.09 | 3.22 | 3.65 | 3.9 | 5.06 | 3.31 | 1.97 |
| $E_{40}$ | 40 років | 6.49 | 2.25 | 2.26 | 2.6 | 2.75 | 3.53 | 2.4 | 1.41 |
| $E_{50}$ | 50 років | 4.16 | 1.51 | 1.51 | 1.76 | 1.8 | 2.33 | 1.63 | 0.93 |
| $E_{60}$ | 60 років | 2.46 | 0.93 | 0.95 | 1.09 | 1.09 | 1.46 | 1.04 | 0.57 |
| $E_{70}$ | 70 років | 1.34 | 0.53 | 0.58 | 0.64 | 0.62 | 0.88 | 0.62 | 0.33 |
| $E_{80}$ | 80 років | 0.67 | 0.28 | 0.33 | 0.35 | 0.32 | 0.51 | 0.35 | 0.18 |
| $E_{90}$ | 90 років | 0.31 | 0.14 | 0.18 | 0.19 | 0.16 | 0.28 | 0.18 | 0.09 |
| $E_{100}$ | 100 років | 0.13 | 0.06 | 0.09 | 0.09 | 0.07 | 0.15 | 0.09 | 0.04 |
| | Очікувана тривалість життя при народженні (роки) | 65.6 | 78.9 | 78.6 | 79.6 | 78.8 | 77.7 | 80.6 | 77.5 |